\pdfoutput=1
\documentclass[aps,preprint,onecolumn,superscriptaddress]{revtex4-1}

\usepackage{graphicx} 
\usepackage{epsfig}
\usepackage{amsmath} 
\usepackage{amsthm} 
\usepackage{amssymb}	
\usepackage{physics}
\usepackage{graphics} 
\usepackage{hyperref} 
\hypersetup{
    colorlinks,
    citecolor=blue,
    filecolor=black,
    linkcolor=red,
    urlcolor=blue
}

\usepackage{bm}
\usepackage{epstopdf}
\usepackage{array}
\usepackage{enumitem}
\usepackage{verbatim} 
\usepackage{epsfig}
\usepackage{tikz}
\usepackage{float}
\usetikzlibrary{shapes, arrows}
\begin{document}

\title{Designing allostery-inspired response in mechanical networks}
\author{Jason W. Rocks}
\affiliation{Department of Physics and Astronomy, University of Pennsylvania, Philadelphia, PA, USA}
\author{Nidhi Pashine}
\affiliation{Department of Physics, University of Chicago, Chicago, IL, USA}
\author{Irmgard Bischofberger}
\affiliation{Department of Mechanical Engineering, Massachusetts Institute of Technology, Cambridge, MA, USA}
\author{Carl P. Goodrich}
\affiliation{School of Engineering and Applied Sciences, Harvard University, Cambridge, MA, USA}
\author{Andrea J. Liu}
\affiliation{Department of Physics and Astronomy, University of Pennsylvania, Philadelphia, PA, USA}
\author{Sidney R. Nagel}
\affiliation{Department of Physics, University of Chicago, Chicago, IL, USA}

\begin{abstract}
Recent advances in designing meta-materials have demonstrated that global mechanical properties of disordered spring networks can be tuned by selectively modifying only a small subset of bonds. Here, using a computationally-efficient approach, we extend this idea in order to tune more general properties of networks. With nearly complete success, we are able to produce a strain between any pair of target nodes in a network in response to an applied source strain on any other pair of nodes by removing only $\sim$1\% of the bonds.  We are also able to control multiple pairs of target nodes, each with a different individual response, from a single source, and to tune multiple independent source/target responses simultaneously into a network. We have fabricated physical networks in macroscopic two- and three-dimensional systems that exhibit these responses.  This targeted behavior is reminiscent of the long-range coupled conformational changes that often occur during allostery in proteins.  The ease with which we create these responses may give insight into why allostery is a common means for the regulation of activity in biological molecules. 
\end{abstract}
\maketitle

\section*{Introduction}
The ability to tune the response of mechanical networks has significant applications for designing meta-materials with unique properties. For example, the ratio $G/B$ of the shear modulus $G$ to the bulk modulus $B$ can be tuned by over 16 orders of magnitude by removing only 2\% of the bonds in an ideal spring network~\cite{Goodrich2015a}.  Such a pruning procedure allows one to create a network that has a Poisson ratio $\nu$ anywhere between the auxetic limit ($\nu=-1$) and the incompressible limit ($\nu= 1/(d-1)$ in $d$ dimensions).  In another example, the average coordination number of a network controls the width of a failure zone under compression or extension~\cite{Driscoll2016}.  Both these results are specific to tuning the global responses of a material. However, many applications rely on targeting a local response to a local perturbation applied some distance away.  For example, allostery in a protein is the process by which a molecule binding locally to one site 
affects the activity at a second distant site~\cite{Ribeiro2016}.  Often this process involves the coupling of conformational changes between the two sites~\cite{Daily2007}. Here we ask whether disordered networks, which generically do not exhibit this 
behavior, can be tuned to develop a specific allosteric structural response by pruning bonds.  

We introduce a formalism for calculating how each bond contributes to the mechanical response anywhere in the network to an arbitrary applied source strain.  This allows us to develop algorithms to control how the strain between an arbitrarily chosen pair of target nodes responds to the strain applied between an arbitrary pair of source nodes. 
In the simplest case, bonds are removed sequentially until the desired target strain is reached.  For almost all of the initial networks studied, only a small fraction of the bonds need to be removed in order to achieve success.  As was the case in tuning the bulk and shear moduli, we can achieve the desired response in a number of ways by pruning different 
bonds.   We have extended our approach to manipulate multiple targets simultaneously from a single source, as well as to create independent responses to different locally applied strains in the same network. 


We demonstrate the success of this method by reproducing our theoretical networks in macroscopic physical systems constructed in two dimensions by laser cutting a planar sheet and in three dimensions by using 3D printing technology.  Thus, we have created a new class of mechanical meta-materials with specific allosteric functions.

Our central result is the ease and precision with which allosteric conformational responses can be created with only minimal changes to the network structure.
This finding can be viewed as a first step towards understanding why allosteric behavior is so common in biopolymers~\cite{Gunasekaran2004}.  It has been emphasized that the ability to control allosteric responses in folded proteins could lead to significant advances in drug design~\cite{Nussinov2013, Guarnera2016}. While much work has focused on identifying, understanding and controlling pre-existing allosteric properties, the question of how to introduce new allosteric functions is relatively unexplored~\cite{Dokholyan2016}. 

\section*{Theoretical Approach}
Our networks are generated from random configurations of soft spheres in three dimensions or discs in two dimensions with periodic boundary conditions that have been brought to a local energy minimum using standard jamming algorithms~\cite{OHern2003, Liu2010}; the spheres overlap and are in mechanical equilibrium.  We convert a jammed packing into a spring network by joining the centers of each pair of overlapping particles with an unstretched central-force spring. We chose this ensemble because it is disordered and provides initial networks with properties -- such as elastic moduli -- that depend on the coordination of the network in ways that are understood~\cite{Ellenbroek2009,Ellenbroek2015, Goodrich2015a}.  We can work either with the entire system that is periodically continued in space or with a finite region with free boundaries that is cut from the initial network.
  
Starting with a network with $N$ nodes and $N_b$ bonds in $d$ dimensions, our aim is to tune the strain $\varepsilon_T$ between a pair of target nodes in response to the strain $\varepsilon_S$ applied between two source nodes. (The two nodes comprising each of the target or source are chosen so that they are not initially connected by a bond; see SI). We create a specific response in our system by tuning the strain ratio $\eta = \varepsilon_T / \varepsilon_S$ to a desired value $\eta^*$.  At each step, we calculate how $\eta$ would change in response to the removal of each bond in the network and then remove the bond which minimizes the difference between $\eta$ and $\eta^*$.

We define the response of the network as a vector of bond extensions $\ket*{e}$ of length $N_b$. We use the orthonormal set of bond basis vectors $\ket*{i}$ to access the extensions of particular bonds. The extension of bond $i$ is attained by $e_i = \bra*{i}\ket*{e}$, while the strain on the bond is then $\varepsilon_i=e_i/l_i$ where $l_i$ is the bond's equilibrium length. The equilibrium matrix $Q$ relates the bond extensions to the vector of node displacements $\ket*{u}$ of length $dN$ through the relation $Q^T\ket*{u} = \ket*{e}$~\cite{Calladine1978}. Additionally, we define the $N_b$-vector of bond tensions $\ket*{t}$ where the tension on bond $i$ is $t_i = \bra*{i}\ket*{t}$. The equilibrium matrix also relates $\ket*{t}$ to the $dN$-vector of net forces on the nodes $\ket*{f}$ by $Q\ket*{t} = \ket*{f}$.

The matrix $Q$ encodes the structure of the network and can therefore be used to define a convenient basis for bond tensions. We find this basis by performing a singular value decomposition of $Q$ and calculating its right singular vectors~\cite{Pellegrino1993}. Our result provides a complete basis of size $N_b$ which is composed of two mutually orthonormal sub-bases. The first sub-basis corresponds to the singular values of $Q$ that are zero; each of these vectors contains a set of tensions on the bonds that do not result in net forces on the nodes. These are commonly known as the \textit{states of self-stress} (SSS). The second sub-basis corresponds to the positive singular values of $Q$; these vectors contain tensions that correspond to net forces on nodes.  We call these the \textit{states of compatible stress} (SCS). 
Using this basis, we can calculate \textit{analytically} the change in bond extensions when a bond is removed (see Methods). This gives us a significant advantage over methods that typically require solving a system of $dN$ equations at each iteration. 

\section*{Computational Results}
We apply our tuning approach to networks with free boundaries in both two and three dimensions (see Methods). We characterize the connectivity of our networks by the excess coordination number $\Delta Z \equiv Z - Z_{\text{iso}}$. Here $Z$ is the average number of bonds per node and $Z_{\text{iso}}\equiv 2d - d(d+1)/N$ is the minimum number of bonds needed for rigidity in a network with free boundary conditions~\cite{Goodrich2012}. For each trial, a pair of source nodes was chosen randomly on the network's surface, along with a pair of target nodes located on the surface at the opposing pole.  (Note that we could have chosen anywhere in the network for the location of the source and target.) 
In two dimensions we chose networks that on average had $190$ nodes and $400$ bonds before tuning, with $\Delta Z \approx 0.19$. In three dimensions networks had on average $240$ nodes, $740$ bonds and $\Delta Z \approx 0.18$. Prior to pruning, the average strain ratio of the networks in two dimensions was $\eta \approx 0.03$ and in three dimensions was $\eta \approx 0.2$ for the systems sizes and $\Delta Z$ values we studied. The response of each network was tuned by sequentially removing bonds until the difference between the actual and desired strain ratios, $\eta$ and $\eta^*$ respectively, was less than 1\%.   

\begin{figure}
	\centering
	\includegraphics[width=0.75\linewidth]{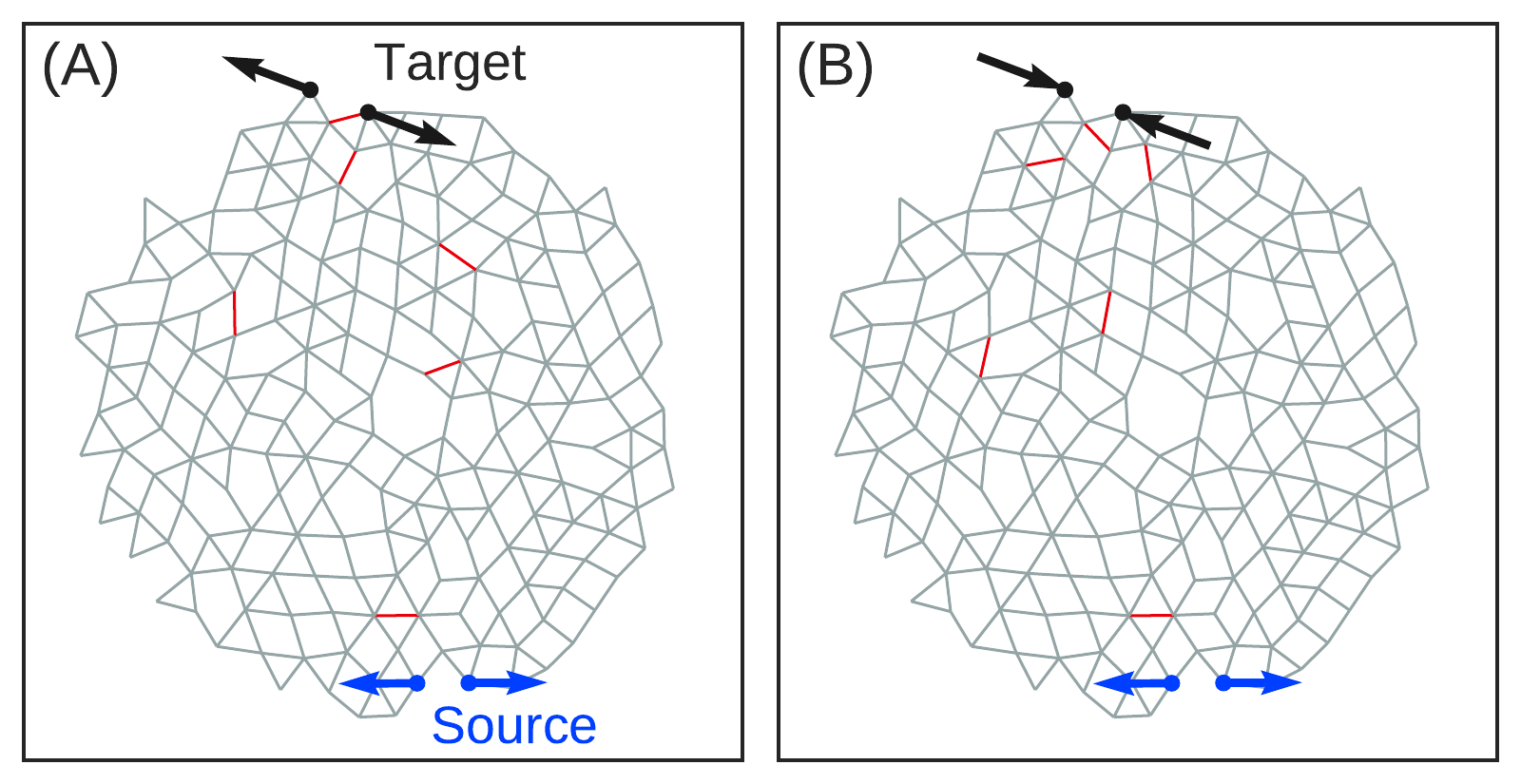}
	\caption{Network with $194$ nodes, $407$ bonds at $\Delta Z = 0.19$ tuned to exhibit (A) expanding ($\eta = +1$) and (B) contracting ($\eta = -1$) responses to within 1\% of the desired response. Source nodes are shown in blue, while target nodes are shown in black. Arrows indicate the sign and magnitude of the extension between the pairs of source and target nodes. The removed bonds are shown as red lines. }
	\label{fig:networks}
\end{figure}

To demonstrate the ability of our approach to tune the response, we show results for $\eta=\pm 1$.  Note that $\eta>0$ ($<0$) corresponds to a larger (smaller) separation between the target nodes when the source nodes are pulled apart. Figure~\ref{fig:networks} shows a typical result for a two-dimensional network:  in Fig~\ref{fig:networks}(A), the strain ratio has been tuned to $\eta= +1$ with just 6 (out of 407) bonds removed; Figure~\ref{fig:networks}(B) shows the same network tuned to $\eta=-1$ with a different set of 6 removed bonds.  The red lines in each figure indicate the bonds that were pruned.  Animations of the full nonlinear responses of these networks are provided in Videos S1 and S2 of the Supporting Information.  We note that some of the removed bonds are the same for both $\eta= +1$ and $\eta= -1$.

\begin{figure}
	\centering
	\includegraphics[width=0.75\linewidth]{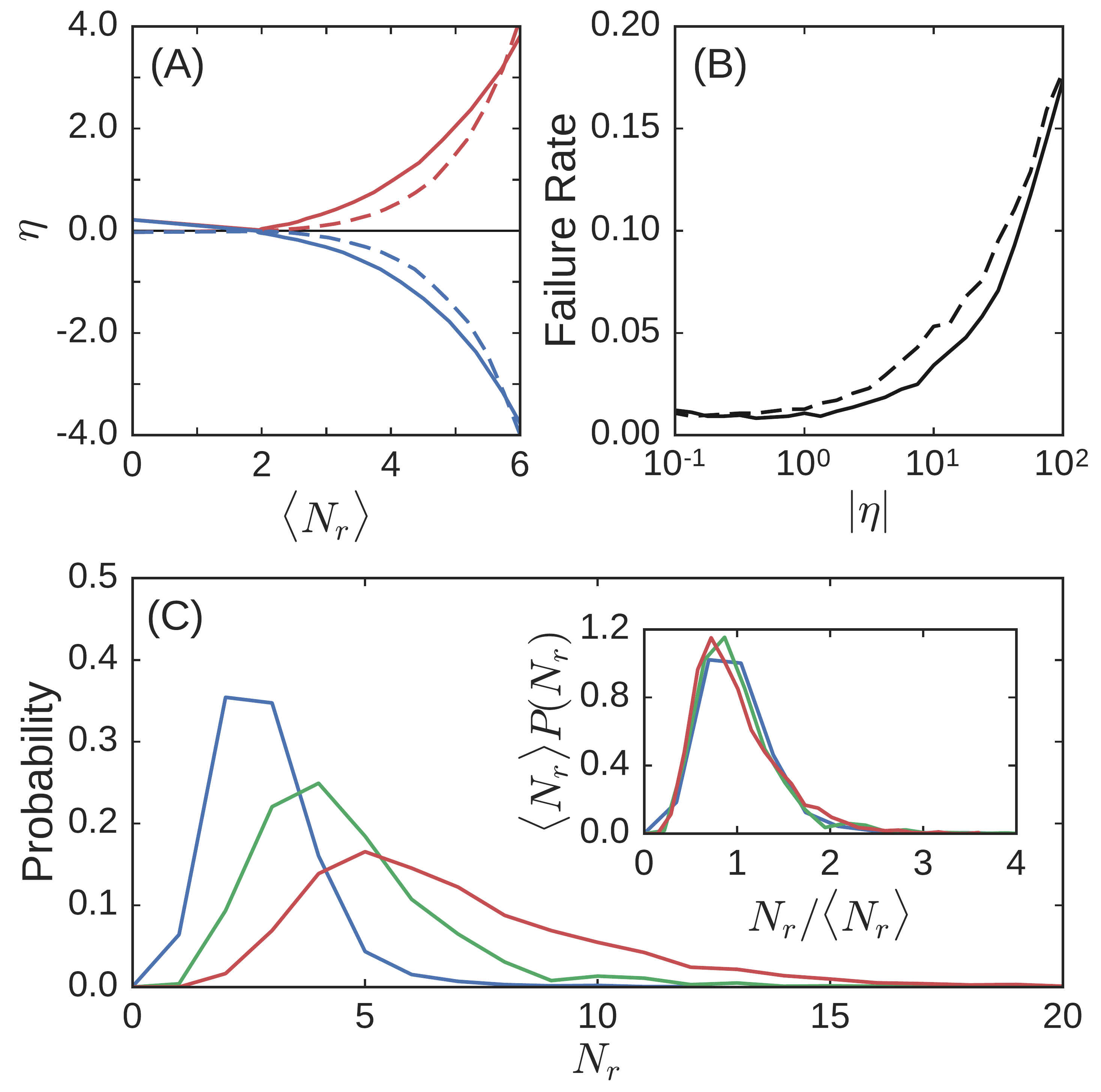}
	\caption{(A) Strain ratio $\eta$ versus the number of removed bonds $N_r$ for expanding (red) and contracting (blue) responses in both 2D (solid lines) and 3D (dashed lines). For each response type and dimension, the strain ratio is averaged over $1024$ tuned networks constructed from $512$ initial systems. Networks in 2D had about $190$ nodes and $400$ bonds on average with an initial excess bond coordination of $\Delta Z \approx 0.19$, while those in 3D had about $240$ nodes and $740$ bonds on average with $\Delta Z \approx 0.18$. (B) Failure rate of tuning systems to within 1\% of a specified strain ratio magnitude in 2D (dashed lines) and 3D (solid lines) averaged over contracting and expanding responses. (C) The distribution of the number of removed bonds for three different strain ratio magnitudes: $|\eta|=0.1$ (blue), $|\eta|=1.0$ (green), and $|\eta|=10.0$ (red). Inset: All three distributions collapse when scaled by the average number of removed bonds $\langle N_r \rangle$. }
	\label{fig:strain_ratio}
\end{figure}

The average strain ratio versus the number of removed bonds is shown in Fig.~\ref{fig:strain_ratio}(A). Remarkably few bonds need to be removed in order to achieve strain ratios of $\eta = \pm1$. In two dimensions only about 5 bonds out of about 400 were removed on average ($\sim$1\%); similarly, in three dimensions only about 4 bonds out of about 740 were removed on average ($\sim$0.5\%). Fig.~\ref{fig:strain_ratio}(B) shows the fraction of networks that cannot be tuned successfully to within 1\% of a given strain ratio. The failure rate is less than 2\% for strain ratios of up to $|\eta|= 1$ in two dimensions and less than 1\% in three dimensions. Therefore, not only does our algorithm allow for precise control of the response, it also works the vast majority of the time. The failure rate increases significantly for $|\eta| \gg 1$, but here we are considering only the linear response of the network. Extremely large values of $\eta$ necessitate an extremely small input strain at the source and may therefore not be physically relevant. 

The failure rate is insensitive to $\Delta Z$ except at very small values. In the small $\Delta Z$ regime the failure rate is higher because very few bonds can be removed without compromising the rigidity of the system. If we increase the bond connectivity to $\Delta Z\approx 1.0$ for networks in two dimensions, the failure rate remains very low, but bonds are removed in a thin region connecting the target and source.  This narrowing of the ``damage'' region is reminiscent of the results of Ref.~\cite{Driscoll2016}, in which bonds above a threshold stress were broken, or of Ref.~\cite{Goodrich2015a}, in which bonds that contribute the most to either the bulk or shear modulus were successively pruned.  

Figure~\ref{fig:strain_ratio}(C) shows the distribution of the number of bonds that must be removed to tune a network to within 1\% of a desired strain ratio for $|\eta|=0.1$, $1$, and $10$.  These distributions are broad and the mean shifts upwards as $\eta$ increases.  The inset shows that the distributions collapse when normalized by the average number of removed bonds $\langle N_r \rangle$. Note that we do not achieve the desired strain ratio simply by tuning the entire free surface of the network to have large strain ratios; the response between the designated target is large while the response between other pairs of nodes is essentially unaffected by the source strain (see Figure S1 in the SI).

\begin{figure}
	\centering
	\includegraphics[width=0.75\linewidth]{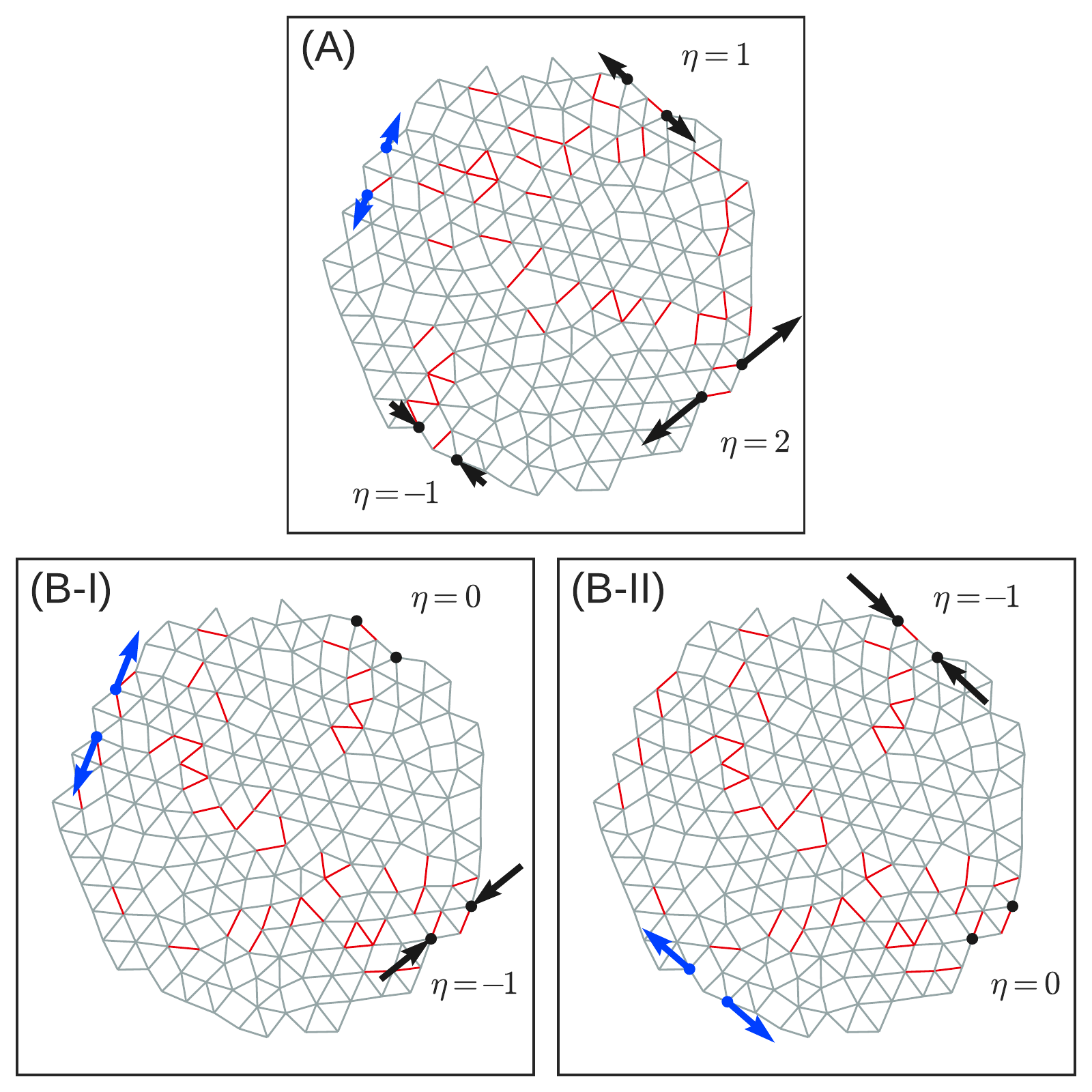}
	\caption{Network with $200$ nodes and $502$ bonds at $\Delta Z = 1.0$. (A) Network tuned to show responses at three targets with responses of $\eta =1$, $2$, and $-1$. All three targets are controlled by a single pair of source nodes. (B) Same network with two independent responses tuned simultaneously into the system. (B-I) One target contracts in response to a strain at the first pair of source nodes while the other target does not respond. (B-II) Second target responds to a strain at the second source while the first target remains unaffected. This demonstrates that separate responses can be shielded effectively from one another.}
	\label{fig:multiout}
\end{figure}

Figure~\ref{fig:multiout} shows we are also able to create systems with more complicated responses. In Fig.~\ref{fig:multiout}(A), a network with one pair of source nodes controls three targets, each of which has been tuned to a different strain ratio. Figures~\ref{fig:multiout}(B-I) and (B-II) show a single network with two independent sources and targets whose responses were tuned simultaneously and independently of one another.  When a strain is applied to the first pair of source nodes, its target responds strongly while the other target does not respond at all. Likewise, when the strain is applied to the second pair of source nodes, its target responds while the first target does not.  
These networks have $\Delta Z = 1.0$; the failure rate for creating these more complicated responses is generally higher for lower values of $\Delta Z$ in two dimensions.  

\section*{Experimental Results}
\begin{figure}
	\centering
	\includegraphics[width=0.75\linewidth]{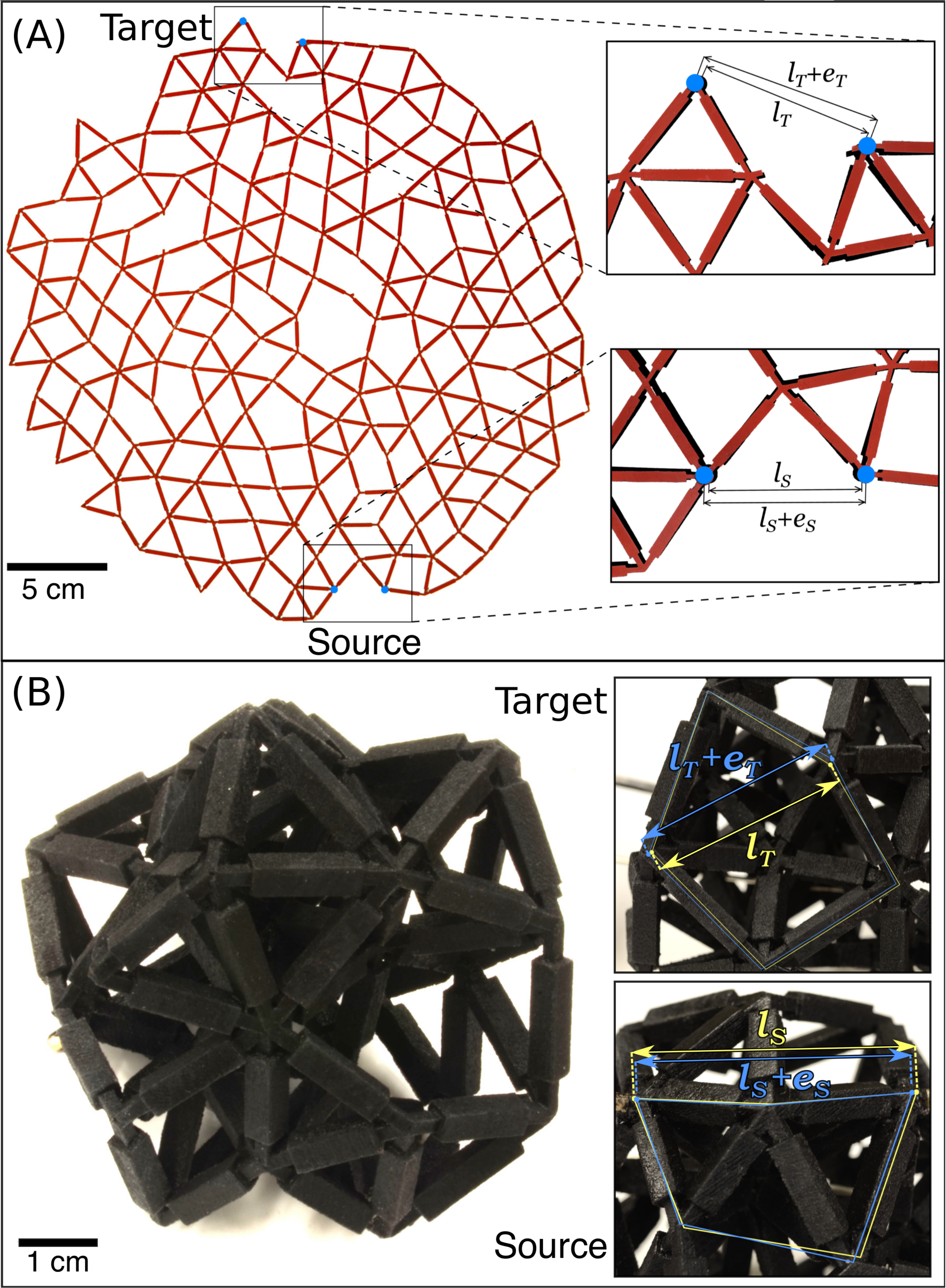}
	\caption{(A) Physical realization of the network in Fig.~\ref{fig:networks}(A).
The zoomed-in photographs show the initial and final distance between the pair of source nodes, $l_S$ and $l_S+e_S$, respectively, and between the pair of target nodes, $l_T$ and $l_T+e_T$. The undeformed network is shown in black, while the deformed network is superimposed in red.
(B) Photograph of a three-dimensional network constructed by 3D printing with 33 nodes and 106 bonds at $\Delta Z = 0.42$ tuned to exhibit a negative response ($\eta = -5.0$).
In the zoomed-in photographs, the yellow and blue arrows show respectively the distance between the undeformed, $l_S$ ($l_T$), and deformed, $l_S+e_S$ ($l_T+e_T$), source (target) nodes.}
	\label{fig:exp}
\end{figure}

Figure~\ref{fig:exp}(A) shows an image of a two-dimensional network created by laser cutting a flat sheet.  The network is the same as the one shown in Fig.~\ref{fig:networks}(A). The zoomed-in areas show the strain response at the target nodes along with the applied strain at the source nodes.  The measured value of $\eta = 1.00\pm 0.01$ is consistent with the theoretical prediction.  Figure~\ref{fig:exp}(B) shows an image of a three-dimensional network created by 3D printing.  In this case, the network was designed to have a strain ratio of $\eta=-5$. The insets again show the relative strains between the pairs of target and source nodes.

In order to obtain a quantitative analysis of how well the physical realizations agree with the simulated networks, we measure the strain on every bond in the two-dimensional example when the distance between the source nodes is varied.  A majority of the bonds do not change their length appreciably.  We therefore focus only on the distance between nodes that were connected by bonds (labeled $i$) that were removed as the network was tuned.  As one might expect, these are the most sensitive to the applied source strain.  We calculate, for those changes in distances, the Pearson correlation coefficient between the experiments and the simulations:
\begin{align*}
C = \frac{\langle(x_i - \langle x_i \rangle)(c_i - \langle c_i \rangle) \rangle}{\sigma_x \sigma_c}
\end{align*}
Here $x_i$ ($c_i$) is defined as the fractional change due to the source strain in the distance between nodes initially connected by bond $i$ as measured in experiments (computer simulations). The standard deviations of $x_i$ and $c_i$ are $\sigma_x$ and $\sigma_c$, respectively. We find that when averaged over 4 experimental realizations of different designed networks, $C = 0.98\pm 0.02$. This indicates that the experiments are a very accurate realization of the theoretical models.

In contrast to our theoretical models, our experimental systems have bond-bending forces that tend to restore angles between bonds to their preferred values. There is also a possibility of buckling out of the plane of two-dimensional networks, along with nonlinear effects that are present in real systems undergoing finite strains. In spite of these differences, experiments show that the responses generated by our linear response theory survive into the nonlinear regime probed by the experiments in a significant fraction of the realizations studied. 
 
\section*{Discussion}
We have shown that it is strikingly easy to tune allosteric deformation responses into an arbitrary spring network by removing only a small fraction of the bonds. Not only can we tune the strain ratio to large negative or positive values for the same network, but we achieve strain ratios of order $|\eta|\sim 1$ with almost 100\% success. Our theoretical approach can also be extended to more general responses. We can control multiple pairs of target nodes simultaneously with the same pair of source nodes and we can tune multiple independent source/target responses simultaneously into a network. We have also achieved similarly excellent results for tuning responses in periodically-continued systems.

The approach we have described here performs a discrete optimization of the response. We have also tuned the response using a standard numerical optimization technique (e.g., gradient descent), by varying the stiffnesses of all the bonds continuously. This brute-force method is less efficient but equally successful in producing a desired response, and has the advantage of being able to tune nonlinear behavior. Our approach can also be generalized to other types of bond manipulation such as introducing new bonds.
 
Our theoretical approach provides a framework for understanding and controlling the response of networks relevant to a wide range of fields.  For example, networks with built-in localized, long-distance responses could be a novel way of designing architectural structures based on disordered frameworks that have added functionalities. In addition, our theoretical approach can be generalized to other problems such as origami, where one may wish to tune the fold structure so that the system folds in a specific way in response to locally applied external forces~\cite{Fuchi2015}. This problem is similar to ours, except that folds are added instead of bonds being removed.  Ref.~\cite{Fuchi2015} introduces an optimization technique in which fold rigidities vary continuously.  This technique is computationally expensive because the network response must be recalculated with each optimization step. 
A generalization of our theoretical approach to origami, using language similar to that of Ref.~\cite{SchenkMarkandGuest2011}, could lead to a more efficient algorithm.

The network responses we create are reminiscent of the localized, long-range-correlated deformations which characterize allostery in proteins.  In fact, folded proteins have long been modeled as elastic networks~\cite{Bahar2010} and the response to localized forces in the resulting networks has been studied~\cite{Atilgan2010}.  Our results demonstrate the ease with which allosteric conformational changes in networks can be achieved by removing a very small set of bonds. As such, it suggests why allostery is so common in large biological molecules. 

Similarly, our finding that networks can be tuned to have a variety of different responses may help elucidate multifunctional behavior~\cite{Favia2009} and multiple allosterically interacting sites~\cite{Yuan2015} in proteins. It has also been observed that small changes in a protein's covalent structure can often change its biochemical function~\cite{Jeffery2016}.  One might ask whether our method could be extended to develop a systematic way to determine which intra-protein interactions to modify to create new allosteric functions. Our success in constructing experimental systems in spite of nonlinear and bond-bending effects suggests that results are often robust even outside the simple linear regime. However, proteins are thermal whereas our networks are athermal structures. Statistical fluctuations in the structure of proteins has been shown to play an important role in allosteric functionality~\cite{Tsai2008, Motlagh2014}. It is thus important to investigate how thermal effects can influence the ability to design a desired response. In addition, protein contact networks generally contain pre-stressed bonds, as well as bond-bending and twisting constraints, while our theoretical networks are constructed in the absence of such effects~\cite{Edwards2012, Thorpe2001, Srivastava2012}.


Further work needs to be done to understand why removing specific bonds achieves the desired response. Our method of identifying the elements of the stress basis associated with individual bonds indicates that these stress states are fundamental to this understanding. The dependence on network size and node connectivity also needs to be understood in greater detail. The limits of our algorithm are not yet known, including the number of targets that can be controlled and the number of independent responses that can be tuned for networks of a given size and coordination. To understand experimental systems ranging from proteins to the macroscopic networks we have fabricated, we must extend the theory to include temperature, dynamics, pre-stress, bond bending, and nonlinear effects due to finite strains. Our approach provides a starting point for addressing these issues.

\section*{Materials and Methods}
\subsection*{Computed Networks and Choice of Source and Target Nodes}
To create a finite network, we choose a cut-off radius from the center of our box and remove all bonds that cross that surface. This process often creates zero energy modes at the boundary of our network.  Since we require rigid networks, we remove nodes associated with these modes. We calculate zero modes by performing a spectral decomposition of the dynamical matrix. For each zero mode calculated this way, we identify the node with the largest displacement amplitude and remove it. We then recalculate the zero modes and repeat this process until no zero modes exist. This method of removing zero modes works in any dimension and does not require an arbitrary threshold for whether a node contributes to a zero mode or not. Our final networks are approximately disc-shaped in two dimensions or ball-shaped in three dimensions with $N$ nodes and $N_b$ bonds. 

We choose the pair of source nodes to lie on the exposed surface of the networks. The pair of target nodes is chosen to be on the opposing pole of the network surface. When choosing a pair of nodes, we also ensure that they are not connected by a bond. This done to avoid surface bonds whose tensions do not couple the the rest of the network. However, since our formalism relies on applying tensions and measuring the strains of bonds, we introduce a bond of zero stiffness, called a ``ghost'' bond, between each pair of nodes for convenience (see SI).

\subsection*{Further Details of Theoretical Approach}

Our approach tunes the ratio $\eta=\varepsilon_T/\varepsilon_S$ of the target strain $\varepsilon_T$ to the source strain by removing bonds sequentially, one at a time.  First, we define the cost function which measures the difference between the network's response $\eta$ and the desired response $\eta^*$. This is given by
\begin{align}
\Delta^2 \equiv \sum_n \left\{\begin{array}{c r}
(\eta_n/\eta_n^* -1)^2 & \qif \eta_n^* \neq 0\\
\eta_n^2 & \qif \eta_n^* = 0
\end{array}\right.\label{eq:cost}
\end{align}
where $n$ indexes the targets and their corresponding sources  (For example, $n=1, 3, 2$ in Figs.~\ref{fig:networks}, Fig.~\ref{fig:multiout}(A) and Fig.~\ref{fig:multiout}(B), respectively.) . Target/source pairs may be defined for the same network response, or for separate independent responses for the same network with different applied source strains. With each step, we choose to remove the bond which creates the largest decrease in $\Delta^2$.

To decide which bond to remove, we must calculate how the removal of each bond changes $\eta$. First we define the vectors of bond extensions $\ket*{e}$ and bond tensions $\ket*{t}$ in response to the externally applied strain, each of length $N_b$. In order to access the extensions and tensions on individual bonds, we define the complete orthonormal bond basis $\ket*{i}$ where $i$ indexes the bonds. The extension on bond $i$ can then be found, $e_i = \bra*{i}\ket*{e}$, along with the bond tension, $t_i =\bra*{i}\ket*{t}$. The strain of bond $i$ is $\varepsilon_i = e_i/l_i$ where $l_i$ is the bond's equilibrium length. The tension and extension are related by a form of Hooke's law, 
\begin{align}
\ket*{t} = F^{-1}\ket*{e}\label{eq:hooke}
\end{align}
where the flexibility matrix is defined as $\mel*{i}{F}{j} =  \delta_{ij}/k_i$. Here we choose the stiffness of bond $i$ to be $k_i=\lambda_i/l_i$ where $\lambda_i$ is the bond's material modulus with units of energy per unit length.

In addition to the bond tensions and extensions, we can define the $dN$-vectors of node displacements $\ket*{u}$ and net forces on nodes $\ket*{f}$. The equilibrium matrix $Q$ relates quantities defined on the bonds to those defined on the nodes through the expressions $Q^T \ket*{u} = \ket*{e}$ and $Q\ket*{t} = \ket*{f}$~\cite{Calladine1978}. In general $Q$, is a rectangular matrix with $dN$ rows and $N_b$ columns. The total energy can then be written
\begin{align}
E = \frac{1}{2}\mel{u}{H}{u}\label{eq:energy}
\end{align}
where the Hessian matrix $H = QF^{-1}Q^T$ is a $dN\times dN$ matrix. In the presence of an externally applied set of tensions $\ket*{t^*}$, the minimum energy configuration satisfies
\begin{align}
H\ket*{u} = Q\ket*{t^*}.\label{eq:syseq}
\end{align}
To calculate the change in the displacements if a bond were removed, the naive approach would be to set the stiffness to zero for that bond in the flexibility matrix and to solve this equation. However, performing this matrix inversion to test the removal of each bond can be prohibitively expensive with a computational cost of $\order*{N_b N^3}$, so we have developed a more efficient approach. Note that here we calculate the response to applied tensions, not the strains we need to calculate $\eta$. However, since we are only interested in the ratio of the target strain to the source strain and are working in the linear regime, we do not need to explicitly apply a strain.  

We use the equilibrium matrix $Q$ to define a convenient basis of the bond tensions and extensions. Performing a singular value decomposition of $Q$ gives access to is right singular vectors~\cite{Pellegrino1993}. This yields two mutually orthonormal sub-bases of vectors that together form a complete basis of size $N_b$. The first sub-basis is comprised of vectors with singular values of $Q$ that are zero; that is, tensions that do not result in net forces on the nodes. These are commonly known as the \textit{states of self-stress} (SSS), and we denote them as $\ket*{s_\beta}$ where $\beta$ indicates the particular basis vector. These vectors can also be interpreted as incompatible extensions, or extensions that do not correspond to valid displacements. The second sub-basis is comprised of vectors with positive singular values of $Q$; tensions that correspond to net forces on nodes, or extensions that are compatible with node displacement. We call these vectors the \textit{states of compatible stress} (SCS) and denote them as $\ket*{c_\alpha}$ where $\alpha$ indicates the basis vector. In total there are $N_c$ SCS basis vectors and $N_s$ SSS basis vectors which total to $N_c+N_s = N_b$.

Using the SSS and SCS sub-bases, our goal is to find the change in $\ket*{e}$ when the stiffness of a given bond is modified. We start by decomposing the bond tensions and extensions,
\begin{align}
\ket*{t} &= \sum\limits_\alpha  \ket*{c_\alpha}\bra*{c_\alpha}\ket*{t} + \sum\limits_\beta \ket*{s_\beta}\bra*{s_\beta}\ket*{t}\\
\ket*{e} &= \sum\limits_\alpha  \ket*{c_\alpha}\bra*{c_\alpha}\ket*{e} + \sum\limits_\beta \ket*{s_\beta}\bra*{s_\beta}\ket*{e}
\end{align}
Now suppose we apply some external tension to the bonds, $\ket*{t^*}$. The part of the external tension that projects onto the SCS basis will be balanced by tensions in the bonds, so that $\bra*{c_\alpha}\ket*{t} = \bra*{c_\alpha}\ket*{t^*}$. Additionally, the bond extensions that project onto the incompatible extensions, or SSS basis, should be zero because they are unphysical, $\bra*{s_\beta}\ket*{e} = 0$. Inserting our decompositions of the tension and extension into \eqref{eq:hooke}, we get
\begin{equation}
\sum\limits_\alpha  \ket*{c_\alpha}\bra*{c_\alpha}\ket*{t^*} + \sum\limits_\beta \ket*{s_\beta}\bra*{s_\beta}\ket*{t} = \sum\limits_\alpha F^{-1}\ket*{c_\alpha}\bra*{c_\alpha}\ket*{e}
\end{equation}
If we project this equation onto the SCS vector $\bra*{c_{\alpha'}}$, we get a system of $N_c$ equations,
\begin{align}
\bra*{c_{\alpha'}}\ket*{t^*} &= \sum\limits_\alpha \mel*{c_{\alpha'}} {F^{-1}}{c_\alpha}\bra*{c_\alpha}\ket*{e}\\
&= \sum\limits_\alpha K_{\alpha'\alpha}\bra*{c_\alpha}\ket*{e}
\end{align}
where $K_{\alpha'\alpha} = \mel*{c_{\alpha'}} {F^{-1}}{c_\alpha}$ is an $N_c \times N_c$ square matrix.  If we invert this system of equations to solve for the extensions, we get
\begin{align}
\bra*{c_\alpha}\ket*{e} &= \sum\limits_{\alpha'} K^{-1}_{\alpha\alpha'} \bra*{c_{\alpha'}}\ket*{t^*}
\end{align}
The full extension is then
\begin{align}
\ket*{e} = \sum\limits_\alpha\ket*{c_\alpha}\bra*{c_\alpha}\ket*{e}
= \sum\limits_{\alpha\alpha'}\ket*{c_\alpha} K^{-1}_{\alpha\alpha'} \bra*{c_{\alpha'}}\ket*{t^*}\label{eq:response}
\end{align}
In general, calculating the matrix inverse $K^{-1}_{\alpha\alpha'}$ is computationally intensive since it is a square matrix of size $N_c$. In order to calculate the change in $\ket*{e}$ upon the removal of each individual bond, it would appear necessary to invert this large matrix $N_b$ times with a runtime of $\order*{N_b N_c^3}$. This would not necessarily be more efficient than solving \eqref{eq:syseq}. To improve this, we can convert freely between a system where all the bonds have different stiffnesses and one where all stiffnesses are unity (see SI). Suppose that all bonds are the same stiffness $k$ except for bond $i$ which has stiffness $k_i$. Now we define the unique SCS basis vector:
\begin{align}
\ket*{C_i} &= \sum\limits_\alpha \ket*{c_\alpha}\bra*{c_\alpha}\ket{i}\label{eq:unique}
\end{align}
This SCS is closely related to the unique SSS defined in Ref.~\cite{Sussman2016}. We can now rotate the SCS basis so that one of the SCS vectors is $\ket*{c_\mu} = \ket*{C_i}/\sqrt{\bra*{C_i}\ket*{C_i}}$, making sure to reorthonormalize the rest of the basis with respect to this unique SCS. The benefit of this rotation is that now only the unique SCS contains a nonzero element for bond $i$. The matrix $K_{\alpha'\alpha}$ can then be simplified to 
\begin{align}
K_{\alpha'\alpha} &= \left\{\begin{array}{cc}
k_i  C_i^2 + k \qty(1 - C_i^2) & \qif \alpha=\alpha' = \mu\\
k \delta_{\alpha\alpha'} & \qotherwise
\end{array} \right.
\end{align}
where we have defined $C_i^2 = \bra*{C_i}\ket*{C_i} = \bra*{i}\ket*{C_i}$. The resulting extensions are 
\begin{align}
\ket*{e} &= \frac{\ket*{C_i}\bra*{C_i}\ket*{t^*}}{C_i^2\qty[k_i  C_i^2 + k \qty(1 - C_i^2)]} + \frac{1}{k}\sum\limits_{\alpha\neq \mu} \ket*{c_\alpha}\bra*{c_\alpha}\ket*{t^*}
\end{align}
If bond $i$ is removed, $k_i$ goes from $k$ to $0$ giving us
\begin{align}
\Delta \ket*{e} &=  \ket*{C_i}\frac{\bra*{C_i}\ket*{t^*}}{k \qty(1-C_i^2)}\label{eq:deltae}
\end{align}
From this equation we can calculate the changes in both $\varepsilon_T$ and $\varepsilon_S$ and therefore the change $\eta$. This result can also be derived by inverting \eqref{eq:syseq} and using the Sherman-Morrison formula to calculate the change in the inverse of the Hessian~\cite{Sherman1950}. Note that this calculation does not include the zero stiffnesses of the ghost bonds, which cannot be mapped to unity with the rest of the system. A generalization of \eqref{eq:deltae} is needed in order to take this into account (see SI).

The next step is to calculate \eqref{eq:deltae} (or its generalization found in the SI) for the removal of each bond. We choose the bond which minimizes $\Delta^2$ in \eqref{eq:cost} upon removal. One restriction is that we do not choose bonds which introduce zero modes (see SI). Finally, once a bond is chosen, we recalculate the SCS and SSS sub-bases with the bond removed (see SI).

A summary of our tuning algorithm contains the following steps:
\begin{enumerate}
\item Transform to a system where all bonds initially have the same stiffnesses and add a ghost bond of zero stiffness for each pair of target and source nodes.
\item Use the equilibrium matrix to calculate the initial SCS and SSS bases.
\item Calculate the initial extensions of the source and target bonds in response to the applied tension $t^*$ using \eqref{eq:response}. Note that initially $K^{-1}_{\alpha\alpha'} = \delta_{\alpha\alpha'}/k$. Use this result to calculate the initial $\eta$. 
\item For each bond, use the general form of \eqref{eq:deltae} found in Eq. SX of the SI to calculate the change in $\eta$ if that bond were to be removed. 
\item Remove the bond that minimizes $\Delta^2$ in \eqref{eq:cost}. Recalculate the SCS and SSS sub-bases with the bond removed.
\end{enumerate}
We repeat (3) - (5) until $\sqrt{\Delta^2}<0.01$ or the process fails.  

There are three potential sources of failure represented in Fig.~\ref{fig:strain_ratio}(B): $\sqrt{\Delta^2}$ cannot be lowered below $0.01$ by removing any bond, no bonds can be removed without creating zero modes, or the numerical error in $\Delta^2$ exceeds $0.01$.  This third source of failure arises because numerical error is  introduced as bonds are removed.  In order to ensure that our results are accurate, we compare $\Delta^2$ to the value obtained from the solution of \eqref{eq:syseq} with the given set of pruned bonds removed.  If the absolute value of the difference exceeds 1\%, we call it a failure.  Our results constitute an upper bound on the failure rate, which could potentially be reduced by using more accurate techniques to decrease numerical error or more sophisticated minimization algorithms.

\subsection*{Experimental Networks}
We create experimental realizations of the theoretically-designed networks in both two and three dimensions. These networks consist of physical struts connecting nodes. Inevitably, there are forces that tend to restore angles between bonds to their preferred values and therefore resist any rotation of the bonds at the nodes. We do not include such bond-bending forces in our calculations, but we design our networks to minimize such effects.

To make two-dimensional networks, we obtain the positions of the nodes and struts from our design algorithm. Next, we laser cut the shape of the network from a silicone rubber sheet. To reduce out-of-plane buckling, we use 1.6 mm thick polysiloxane sheets with a Shore value of A90. The ratio of strut length to width within the plane of the network is approximately 10:1. To minimize the forces due to bond-bending, the struts are made half as wide at their ends - close to each node - as they are at their centers (see Fig.~\ref{fig:exp}(A)). This difference in width at different points ensures that the struts will deform primarily near the nodes rather than buckling at their centers.  

To make three-dimensional networks, we determine the positions of nodes and struts from the computer simulations and fabricate the networks using 3D printing technology.  The proprietary material is a mixture of rubber (simulating styrene based thermoplastic elastomers) and rigid plastic (simulating acrylonitrile butadiene styrene, ABS) with a Shore value of A85.  The dimensions of each strut have a ratio of approximately 1:1:11.  As in our two-dimensional networks, the struts are made thinner at their ends in order to alleviate bond-bending.     

\section*{Acknowledgements}
We thank S. Leibler, T. Tlusty and M. Mitchell for discussions that inspired this work, and the Simons Center for Theoretical Biology at the Institute for Advanced Study for its hospitality to AJL. We also thank D. Hexner, A. Murugan, J. Onuchic and D. Thirumalai for instructive discussions.  This research was supported by the US Department of Energy, Office of Basic Energy Sciences, Division of Materials Sciences and Engineering under Awards DE-FG02-05ER46199 (AJL) and DE-FG02-03ER46088 (SRN) and by a National Science Foundation Graduate fellowship (JWR). This work was partially supported by grants from the Simons Foundation (305547 and 327939 to AJL) and by the National Institute of Standards and Technology under Award 60NANB15D055 (SRN).

\bibliographystyle{unsrt}
\bibliography{allostery_library.bib}

\clearpage

\clearpage
\section*{Supporting Information}
\subsection*{Ghost bonds}
On the surface of our networks there are many nodes with exactly $d$ bonds in $d$ dimensions. Any bond attached to one of these nodes is uncoupled from the rest of the network - applying a tension to one of these bonds does not communicate any tensions or extensions to the rest of the network to linear order. Likewise, no extensions can be measured on these bonds when a tension is applied elsewhere in the network. Therefore, we avoid choosing pairs of source or target nodes that are connected by uncoupled bonds. This is done by ensuring that neither the  pair of source nor target nodes share a bond.

However, all calculations involve the bonds, so in order to apply a tension or measure an extension between two nodes, it is convenient if they share a bond. To apply our approach, we introduce a ``ghost'' bond of zero stiffness between each pair of source or target nodes. These bonds do not affect our results, but allow us to work without any direct reference to the nodes. 

\subsection*{Creating identical bond stiffnesses}
In order to calculate Eq.~(15) in the main text, it was necessary to work in a system where all bonds had identical stiffness. However, we do not want to be restricted to systems that satisfy this special requirement. The bonds in our experimental systems all have the same material modulus $\lambda_i = \lambda$, but their equilibrium lengths $l_i$ differ, resulting in bonds with non-identical stiffnesses $k_i = \lambda/l_i$.  To handle this, we start with a system in which the bond stiffnesses are all different and map it onto an equivalent system in which all the default bond stiffnesses are identical. This is done by introducing a flexibility matrix $F$ (as defined below Eq.~2 in the main text) and scaling the equilibrium matrix so that $\bar{Q} = QF^{-\frac{1}{2}}$. (Note: We can only scale out stiffnesses that are nonzero.) The energy can then be written in terms of $\bar{Q}$:
\begin{align}
E &= \frac{1}{2} u^T \bar{Q} \bar{F}^{-1} \bar{Q}^T u
\end{align}
where the scaled flexibility matrix $\bar{F}$ is proportional to the identity matrix except for any entries that are zero. This energy is the same as that in Eq.~(3) in the main text, so the minimum energy configurations should have a one-to-one correspondence. Scaled extension or tension vectors are related to the unscaled versions by
\begin{align}
\ket*{\bar{e}} &= F^{-\frac{1}{2}}\ket*{e}\\
\ket*{\bar{t}} &= F^{\frac{1}{2}}\ket*{e}
\end{align}
Thus we have implicitly performed all of our calculations on the scaled system and have converted back when calculating $\eta$.

\subsection*{Modifying multiple bonds}
Here we extend Eq.~(15) to allow for multiple bonds that do not have identical stiffness $k$. Suppose that all the bond stiffnesses are identically $k_i=k$, except for a small subset of bonds which we call $\mathcal{B}$. We say a bond $i\in \mathcal{B}$ if and only if $k_i \neq k$. This set includes any ghost bonds with zero stiffness, along with bonds that are being tested for removal or modification. As discussed in the main text, we typically include just three bonds in $\mathcal{B}$ -- the source and target ghost bonds of zero-stiffness, along with the bond tested for removal.

Our goal now is to rotate our SCS sub-basis $\ket*{c_\alpha}$ so that as few basis vectors as possible project onto the bonds in $\mathcal{B}$. We will denote this new rotated SCS sub-basis $\ket*{\tilde{c}_\alpha}$. We define a special set of basis vectors $\mathcal{V}$ such that $\tilde{c}_\alpha\in \mathcal{V}$ if and only if $\bra*{i}\ket*{\tilde{c}_\alpha} \neq 0$ for some $i\in \mathcal{B}$. Typically, the size of $\mathcal{B}$ will equal the size of $\mathcal{V}$.

In order to calculate our rotated SCS sub-basis, $\ket*{\tilde{c}_\alpha}$, we first find the unique SCS for each bond in $\mathcal{B}$, which we denote $\ket*{C_i}$ shown in Eq.~(12). These unique SCS vectors then are orthonormalized using a modified Graham-Schmidt algorithm. The result is the set of basis vectors $\mathcal{V}$ described previously. The remainder of the rotated SCS basis is found by using the modified Graham-Schmidt algorithm to orthonormalize the original SCS basis with respect to the set of vectors $\mathcal{V}$, throwing out any vectors that are completely zeroed out. The result is our set of $N_c$ orthonormal rotated SCS vectors. However, it will be shown that only the vectors in $\mathcal{V}$ will be necessary for our solution.  

Each basis vector that is not in $\mathcal{V}$ has zero projection onto bonds that are in $\mathcal{B}$, i.e. if $\alpha \notin \mathcal{V}$, then $\bra*{i}\ket*{\tilde{c}_\alpha}=0$ for all $i\in\mathcal{B}$. This means that if either $\alpha\notin\mathcal{V}$ or $\alpha'\notin\mathcal{V}$, then $\ket*{\tilde{c}_\alpha}$ and $\ket*{\tilde{c}_{\alpha'}}$ are orthogonal over a reduced basis such that 
\begin{align}
\bra*{\tilde{c}_\alpha}\qty(\sum\limits_i\ket*{i}\bra*{i})\ket*{\tilde{c}_{\alpha'}} = \bra*{\tilde{c}_\alpha}\qty(\sum\limits_{i\notin\mathcal{B}}\ket*{i}\bra*{i})\ket*{\tilde{c}_{\alpha'}}
\end{align}
This new basis now gives us the means to rewrite $K_{\alpha\alpha'}$ for $\alpha\notin\mathcal{V}$ or $\alpha'\notin\mathcal{V}$,
\begin{align}
K_{\alpha\alpha'} &= \mel*{c_{\alpha'}}{F^{-1}}{c_\alpha}\\
&= \bra*{\tilde{c}_\alpha}\qty(\sum\limits_{i\in\mathcal{B}}k_i\ket*{i}\bra*{i}+\sum\limits_{i\notin\mathcal{B}}k\ket*{i}\bra*{i})\ket*{\tilde{c}_{\alpha'}}\\
&=k\sum\limits_{i\notin\mathcal{B}}\bra*{\tilde{c}_\alpha}\ket*{i}\bra*{i}\ket*{\tilde{c}_{\alpha'}}\\
&=k\bra*{\tilde{c}_\alpha}\ket*{\tilde{c}_{\alpha'}} =k\delta_{\alpha\alpha'}
\end{align}
The total matrix is then
\begin{align}
K_{\alpha\alpha'}  = \left\{\begin{array}{c l}
\tilde{K}_{\alpha\alpha'}  & \qif \alpha,\alpha'\in\mathcal{V}\\
k \delta_{\alpha\alpha'} & \qotherwise
\end{array} \right.
\end{align}
where we have defined the sub-matrix $\tilde{K}_{\alpha\alpha'} =\mel*{c_{\alpha'}}{F^{-1}}{c_\alpha}$ for $\alpha,\alpha'\in\mathcal{V}$. We see that the matrix inversion problem is simplified to just inverting $\tilde{K}_{\alpha\alpha'}$. 
\begin{align}
K^{-1}_{\alpha\alpha'} &= \left\{\begin{array}{c l}
\tilde{K}^{-1}_{\alpha\alpha'}  & \qif \alpha,\alpha'\in\mathcal{V}\\
\frac{1}{k} \delta_{\alpha\alpha'} & \qotherwise
\end{array} \right.
\end{align}
Since the size of $\mathcal{B}$ is very small (in our case typically a set of size $3$), calculating the inverse of this matrix is very fast. The extension can now be represented as
\begin{align}
\ket*{e} &= \sum\limits_{\alpha,\alpha'\in\mathcal{V}}\ket*{\tilde{c}_\alpha} \tilde{K}^{-1}_{\alpha\alpha'} \bra*{\tilde{c}_{\alpha'}}\ket*{t^*} + \sum\limits_{\alpha\notin\mathcal{V}} \frac{1}{k}\ket*{\tilde{c}_\alpha}\bra*{\tilde{c}_\alpha}\ket*{t^*}
\end{align}
The change in extension on bond $i$ when the stiffnesses are modified is then
\begin{align}
\Delta \ket*{e} &= \sum\limits_{\alpha,\alpha'\in\mathcal{V}}\ket*{\tilde{c}_\alpha} \Delta (\tilde{K}^{-1}_{\alpha\alpha'})\bra*{\tilde{c}_{\alpha'}}\ket*{t^*}
\end{align}
Note that the solution only depends on the basis vectors in $\mathcal{V}$. This means that only this small number vectors must calculated and the rest my be neglected.

\subsection*{Avoiding the introduction of zero modes}
We impose the constraint that we do not introduce any zero energy modes into the system when we remove bonds.  We ensure this by only removing bonds which contribute to the SSS sub-basis. By Maxwell-Calladine counting,  $N_0 - N_s = dN - N_b$, where $N_0$ is the number of zero modes~[13]. This means that if we remove a bond, we can either add a zero mode (increase $N_0$) or remove a SSS (decrease $N_s$). If a bond is removed that contributes to the SSS sub-basis, a unique SSS will also be removed and no zero mode will be created~[28]. This unique SSS, which we define as $\ket*{S_i}$ for bond $i$, is calculated analogously to the unique SCS, $\ket*{C_i}$, shown in Eq.~(12). We find that
\begin{align}
\ket*{S_i} = \sum\limits_\beta \ket*{s_\beta}\bra*{s_\beta}\ket*{i}
\end{align}
As long as $S_i^2  \equiv  \bra*{S_i}\ket*{S_i}> 0$, then bond $i$ contributes to the SSS sub-basis.

\subsection*{Removing bonds from SSS and SCS sub-bases}
When a bond $i$ is removed, we remove the unique SSS vector $\ket*{S_i}$ from our SSS sub-basis by subtracting off its projection onto each SSS basis vector. We also remove the entries for bond $i$ from all vectors in both our SSS and SCS sub-bases. The two bases are then reorthonormalized using a modified Graham-Schmidt algorithm. This procedure dominates the computational complexity of our algorithm with a computational cost of $\order*{N_b^3}$. However, this is still significantly faster than solving Eq.~(4) each time a bond is removed, which has a cost of $\order*{N_bN^3}$.

\subsection*{Animations of nonlinear response}
Videos \ref{vid:exp} and \ref{vid:cont} show animations of the responses of the networks in Fig. 1 in the main text. Although our algorithm only considers and controls the linear response, we show the full nonlinear deformations. The resulting strain ratio in the nonlinear regime is within a factor of two of the desired linear result up to source strains of $\varepsilon_S=40\%$.

To calculate the nonlinear response, we start with a tuned network and minimize the nonlinear configurational energy for increments of the source strain from 0\% to 40\%. In the network's undeformed state, each node $i$ has an initial position vector $\vb{R}_i$. If the network is deformed in some way, then each node will have a new position
\begin{align}
\vb{X}_i=\vb{R}_i + \vb{u}_i
\end{align}
where $\vb{u}_i$ is the node's displacement vector. If two nodes share a bond, then the bond vector going from node $i$ to node $j$ is $\vb{l}$ with magnitude $l_{ij}$, while the deformed bond vector is
\begin{align}
\vb{X}_{ij} &= \vb{X}_j - \vb{X}_i\\
&= \vb{R}_j - \vb{R}_i + \vb{u}_j - \vb{u}_i\\
&= \vb{l}_{ij} + \Delta\vb{u}_{ij}
\end{align}
where $\Delta\vb{u}_{ij}$ is the bond extension vector. The configuration energy summed over all bonds $\langle ij \rangle$ with central-force harmonic potentials is then
\begin{align*}
E &= \sum\limits_{\langle ij \rangle} \frac{1}{2} k_{ij} \qty(X_{ij} - l_{ij})
\end{align*}
where $X_{ij} = |\vb{X}_{ij}|$ and $k_{ij}$ is the stiffness of bond $\langle ij \rangle$. We minimize this energy numerically with respect to the displacement vectors $\vb{u}_i$ under the constraint that the source strain $\varepsilon_S$ is a specified value.

\clearpage

\renewcommand\thefigure{S\arabic{figure}}
\setcounter{figure}{0}

\begin{figure}
	\centering
	\includegraphics[width=0.75\linewidth]{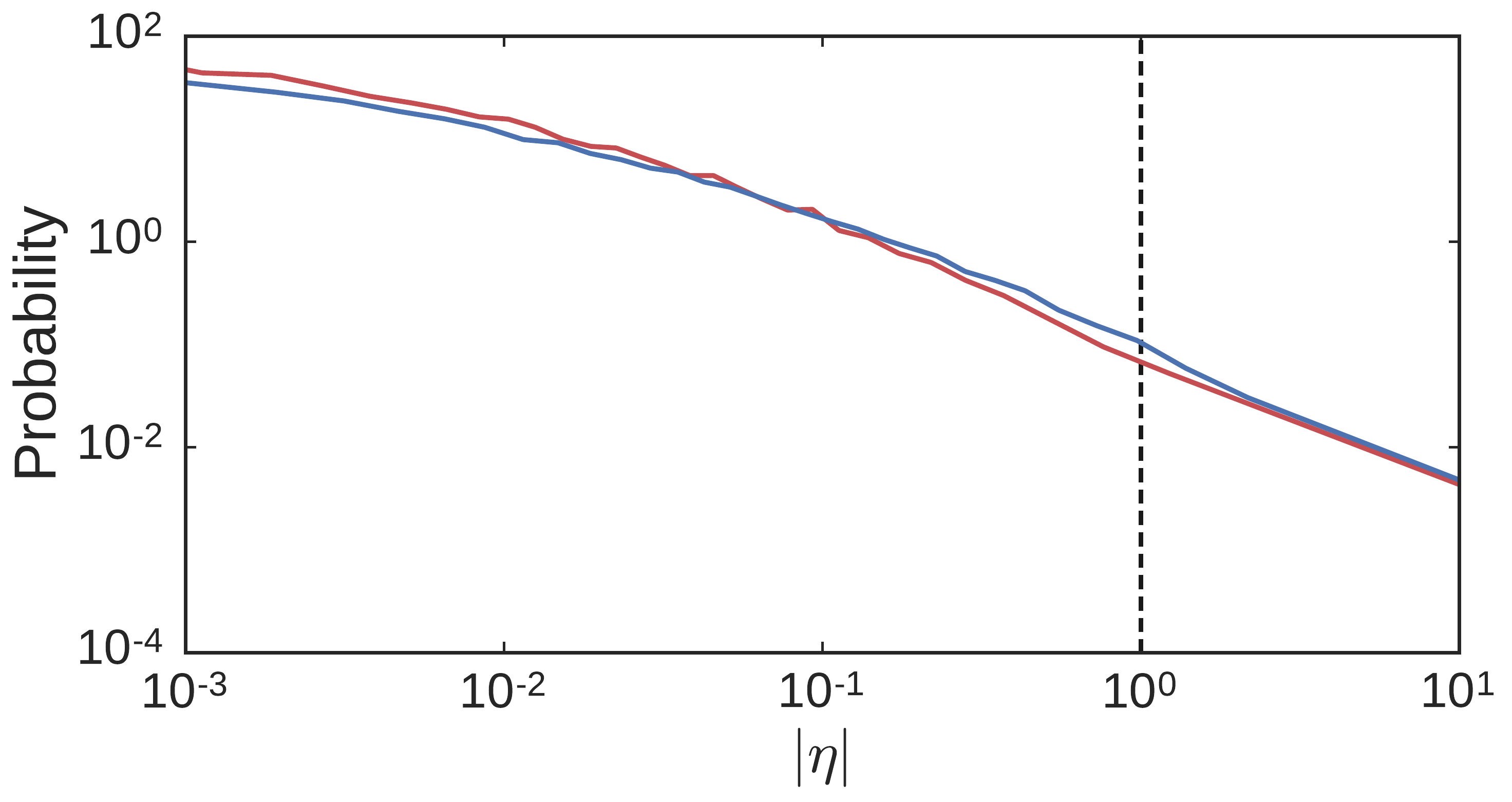}
	\caption{Distribution of strain ratios for pairs of neighboring nodes on the surface (excluding the source and target pairs) of the networks before (red) and after (blue) tuning. The response for pairs other than the source and target are essentially unaffected.  The target strain ratio is shown with a vertical dashed line. All distributions include both contracting and expanding responses. }
	\label{fig:surf_dist}
\end{figure}

\renewcommand{\figurename}{Video }
\renewcommand\thefigure{S\arabic{figure}}
\setcounter{figure}{0}

\begin{figure}
	\centering
	\includegraphics[width=0.4\linewidth]{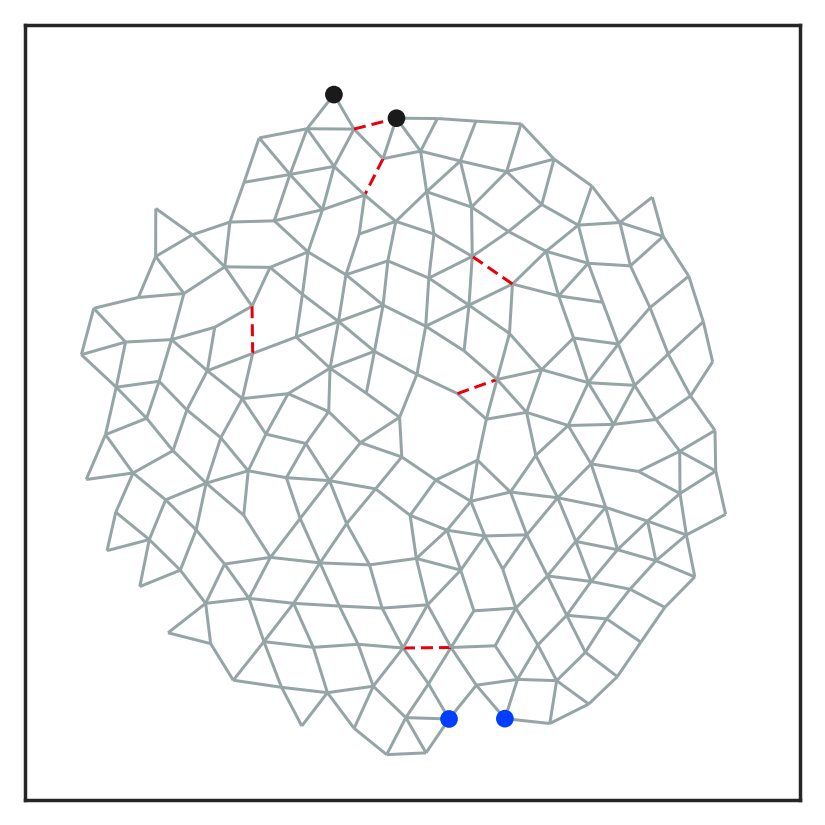}
	\caption{Animation of the tuned expansion response of the network in Fig. 1(A). The source nodes a indicated in blue, while the target nodes are black. The network has been tuned to have a strain ratio of $\eta = +1$ in the linear regime. Here we calculate the full nonlinear response for oscillatory source strain of amplitude 40\% by minimizing the nonlinear configurational energy (see SI). See video at: \url{ https://www.youtube.com/v/iIeeyTBFp6w?playlist=iIeeyTBFp6w&autoplay=1&loop=1}}
	\label{vid:exp}
\end{figure}

\begin{figure}
	\centering
	\includegraphics[width=0.4\linewidth]{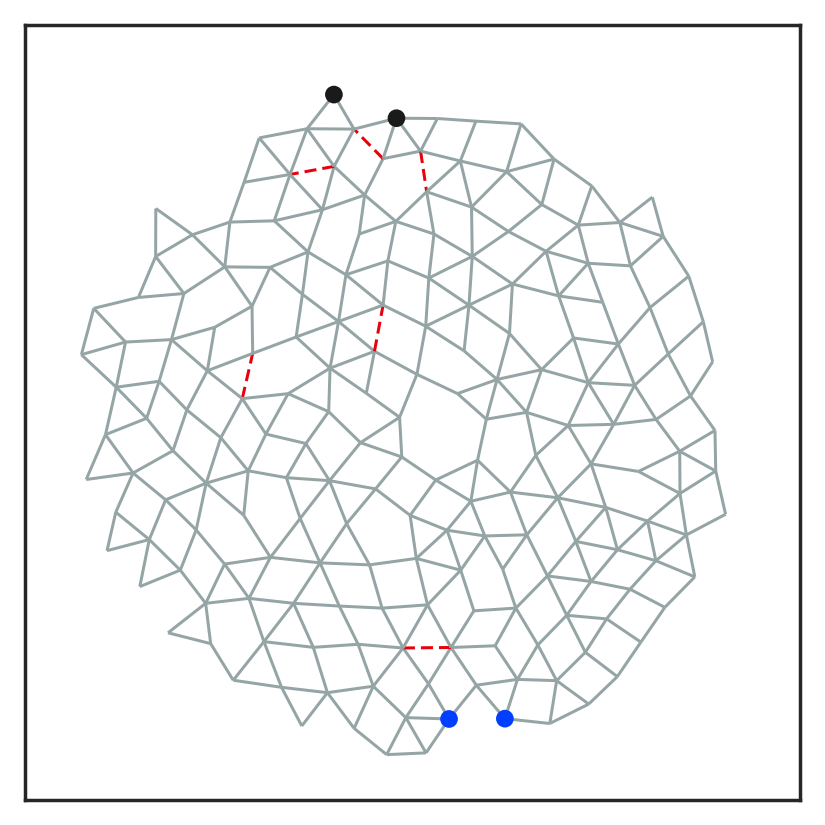}
	\caption{Animation of the tuned contraction response of the network in Fig. 1(B). The source nodes a indicated in blue, while the target nodes are black. The network has been tuned to have a strain ratio of $\eta = +1$ in the linear regime. Here we calculate the full nonlinear response for oscillatory source strain of amplitude 40\% by minimizing the nonlinear configurational energy (see SI). See video at: \url{ https://www.youtube.com/v/drM7JsEtrSA?playlist=drM7JsEtrSA&autoplay=1&loop=1}}
	\label{vid:cont}
\end{figure}
\end{document}